\documentclass[12pt]{article}
\usepackage{epsfig}
\begin{document}
\begin{center} {\bf TEMPERATURE DEPENDENCE OF THE CASIMIR EFFECT}

\vspace{1cm} I. Brevik\footnote{E-mail:  iver.h.brevik@ntnu.no}

\bigskip

Department of Energy and Process Engineering, Norwegian University
of Science and Technology, N-7491 Trondheim, Norway

\bigskip

J. B. Aarseth\footnote{E-mail:  jan.b.aarseth@ntnu.no}

\bigskip
Department of Structural Engineering, Norwegian University of
Science and Technology, N-7491 Trondheim, Norway
\end{center}

\bigskip

\begin{abstract}
In view of the increasing accuracy of Casimir experiments, there
is a need for performing accurate theoretical calculations. Using
accurate experimental data for the permittivities we present, via
 the Lifshitz formula applied to  the standard Casimir setup with
two parallel plates, accurate theoretical results in case of the
metals Au, Cu and Al. Both similar and dissimilar cases are
considered. Concentrating in particular on the finite temperature
effect, we show how the Casimir pressure varies with separation
for three different temperatures, $T=\{1, 300, 350\} $K. The metal
surfaces are taken to be perfectly plane. The experimental data
for the permittivities are generally yielding results that are in
good agreement with those calculated from the Drude relation with
finite relaxation frequency.

We give the results in tabular form, in order to facilitate the
assessment of the temperature correction which is on the 1\%
level. We emphasize two points: (i) The most promising route for a
definite experimental verification of the finite temperature
correction appears to be to concentrate on the case of {\it large}
separations (optimum around $2\,\mu$m); and (ii) there is no
conflict between the present kind of theory and the Nernst theorem
in thermodynamics.
\end{abstract}

\bigskip


\section{Introduction}

Consider the standard Casimir configuration, namely two
semi-infinite homogeneous media separated by a vacuum gap of width
$a$ (recent reviews on the Casimir effect are given in
\cite{milton04,milton01,bordag01,lamoreaux05}). Assume that the
surfaces are perfectly plane, and that they are of infinite
extent.  Denote the left hand slab by 1, the  intermediate vacuum
region by 2, and the right hand slab by 3. The two  permittivities
are $\varepsilon_1(\omega)$ and $\varepsilon_3(\omega)$. Spatial
dispersion is neglected. Working in terms of complex frequencies
$\zeta$, we have $\omega=i\zeta$. The Lifshitz variables $s$ and
$p$, and the Matsubara frequencies $\zeta_m$, are
\begin{equation}
s=\sqrt{\varepsilon-1+p^2}, \quad p=\frac{q}{\zeta_m},\quad
\zeta_m=\frac{2\pi m}{\beta}, \label{1}
\end{equation}
where $\beta=1/T$ is the inverse temperature, and $m$ the
Matsubara integer (we usually put $\hbar =c=1$). The
nondimensional frequency $y$ and the nondimensional temperature
$\gamma$ are defined by
\begin{equation}
y=qa,\quad q=\sqrt{k_\perp^2+\zeta_m^2},\quad \gamma =\frac{2\pi
a}{\beta}, \label{2}
\end{equation}
$\bf{k}_\perp$ being the transverse wave vector (i.e. parallel to
the surfaces). Now defining quantities $\Delta$'s related to the
TE and TM  modes by
\[
\Delta_1^{TE}=\frac{s_1-p}{s_1+p}, \quad
\Delta_2^{TE}=\frac{s_3-p}{s_3+p},\],
\begin{equation}
\Delta_1^{TM}=\frac{\varepsilon_1\,p-s_1}{\varepsilon_1\,p+s_1},
\quad
\Delta_2^{TM}=\frac{\varepsilon_3\,p-s_3}{\varepsilon_3\,p+s_3},
\label{3}
\end{equation}
we can write the Casimir pressure as
\begin{equation}
{\cal F}=-\frac{1}{\pi \beta
a^3}{\sum_{m=0}^\infty}'\int_{m\gamma}^\infty y^2dy \left[
\frac{\Delta_1^{TM}\Delta_2^{TM}e^{-2y}}
{1-\Delta_1^{TM}\Delta_2^{TM}e^{-2y}}+\frac{\Delta_1^{TE}\Delta_2^{TE}e^{-2y}}
{1-\Delta_1^{TE}\Delta_2^{TE}e^{-2y}}\right],\label{4}
\end{equation}
where the prime on the summation sign means that the $m=0$ term is
counted with half weight.

The above formalism summarizes the exposition given recently in
\cite{bentsen05}. The main purpose of the present paper is the
following:

1.  We calculate the Casimir pressure more accurately than we did
earlier, inserting updated permittivity data for the metals Au, Cu
and Al (courtesy of Astrid Lambrecht), and extend also the region
of separation to larger values. The numerical corrections are
small, on the 1\% level, but the generally improving accuracy of
pressure experiments accentuates the need for working to this
degree of accuracy. As before, we take the plates to be perfectly
smooth; roughness corrections have to be dealt with separately and
are not covered here.

A general property of the Casimir pressure is that from $T=0$
onwards it {\it decreases} with increasing values of $T$, given a
fixed value of $a$. However when the separation is large,
$a>3\,\mu$m, the pressure increases with increasing $T$ when $T$
becomes high. Our numerical results indicate that the pressure
increases from around room temperature and upwards. As a general
conclusion, a separation of about $2\,\mu$m appears to be optimal
for testing the temperature correction to the Casimir pressure;
then the relative correction is highest (cf. also figure 5 in
\cite {bentsen05}). The practical problem, of course, is that at
large separations the Casimir pressure itself is small.

2.  We next emphasize the point that there is no conflict between
this kind of theory (which is equivalent to adopting the Drude
dispersion relation) and the Nernst theorem in thermodynamics.
According to this theorem, the entropy at zero temperature has to
be equal to zero. The Nernst theorem is satisfied in our case, in
spite of the fact that the contribution from the TE zero mode
(that is, the $m=0$ term in the Matsubara sum) is equal to zero
for finite $\varepsilon$ as well as for a real metal. The point
here for a real metal is that the relaxation frequency stays
different from zero. The first to emphasize this kind of behaviour
were Bostr{\"o}m and Sernelius \cite{bostrom00}. We have treated
these thermodynamical issues ourselves also
\cite{hoye03,brevik04,brevik05,hoye05}, and there are several
other works expressiong the same opinion
\cite{sernelius04,bostrom04,sernelius05,jancovici05}. We mention,
though, that the opposite view has also been advocated in recent
papers  \cite{bezerra05,decca05,geyer05}.

\section{Calculation and results}

Let us first recall the Drude dispersion relation
\begin{equation}
\varepsilon(i\zeta)=1+\frac{\omega_p^2}{\zeta(\zeta+\nu)},
\label{5}
\end{equation}
where $\omega_p$ is the plasma frequency and $\nu$ is the
relaxation frequency. The plasma wavelength is $\lambda_p=2\pi
c/\omega_p$. or the three metals mentioned, the corrected data as
compared with those given in \cite{bentsen05} are
\[ \omega_p=9.03\,{\rm eV}, \quad \nu=34.5\,{\rm meV},\quad
\lambda_p=137.4\,{\rm nm} \quad {\rm Au},\]
\[ \omega_p=8.97\,{\rm eV},\quad \nu=29.5\,\,{\rm meV},\quad
\lambda_p=138.3\,{\rm nm}\quad {\rm Cu}, \]
\begin{equation}
\omega_p=11.5\,{\rm eV},\quad \nu=50.6\,{\rm meV},\quad
\lambda_p=107.9\,{\rm nm}\quad {\rm Al}. \label{6}
\end{equation}
These corrections are roughly on the 1\% level. As before, we
calculate the Casimir pressure by means of MATLAB, extracting the
zero-frequency case $m=0$ for a separate analytical treatment.
Since $\varepsilon $ becomes very large in the zero frequency
limit for metals, we can express the $m=0$ contribution as
\begin{equation}
{\cal F}_0=\frac{1}{\pi \beta a^3}I_0, \label{7}
\end{equation}
where the polylog function with arguments (3,1) is involved,
\begin{equation}
I_0=-\frac{1}{2}\int_0^\infty y^2dy \frac{ e^{-2y}}{1- e^{-2y}}=
-\frac{1}{8}{\rm polylog}(3,1) = -0.1502571129. \label{8}
\end{equation}
Calculated values of the Casimir pressure for Al, Cu and Al are
shown in tables 1-6, both for the similar and the dissimilar
cases. As in \cite{bentsen05}, we took the temperature $T=1\,$K to
represent the case $T=0$ case with good accuracy. In the
calculations our tolerance for the integrals was $10^{-12}$,
whereas the tolerance  in the sum was $10^{-8}$. At $T=1\,$K the
necessary number of terms was quite large, especially at small
separations (for instance, about 25700  at $a=0.16\,\mu$m).

It is seen that the room-temperature pressure is always weaker
than the zero-temperature pressure. Thus for Au-Au plates at
separation $a=0.5\,\mu$m, the pressure is lowered from 16.56 mPa
to 15.49 mPa, or by 6.5\%. The reduction becomes much enhanced at
larger separations; thus at $a=2\,\mu$m the pressure is lowered
from $7.549\times 10^{-2}$ mPa to $5.550\times 10^{-2}$ mPa, or by
26.5\%. It thus seems advantageous to work with high separations,
if technically possible. The differences between Au-Au and Cu-Cu
pressures are generally small, whereas the pressures for Al-Al are
larger, as we might expect from the dispersive data in (\ref{6}).

As for the cases where dissimilar metals are involved, the Au-Cu
data, table 4, are quite similar to those given in tables 1 and 2.
The data for Au-Al and Cu-Al, tables 5 and 6, show somewhat larger
pressures.

In the tables we show also the pressures when $T=350\,$K, since
one may expect that the pressure difference between 300 K and 350
K will be soon measurable.  Again considering Au-Au at
$a=0.5\,\mu$m, we see that the pressure is lowered from 15.49 mPa
to 15.30 mPa, or by 1.2\%,  when $T$ is increased from 300 K to
350 K. If $a=2\,\mu$m, the corresponding pressure decrease is
3.7\%.

A striking property is that for the larger separations, the
pressure {\it increases} with increasing values of $T$. This turns
out numerically when $a$ becomes larger than about $2.8\,\mu$m.

Finally, the following point should be noted: we did not have to
use the Drude relation in any of our calculations, for any finite
frequency. All the frequencies that were needed, were lying within
the region of Lambrecht's data. We needed the Drude relation
explicitly  at only  one place, namely in the evaluation of the
zero frequency term, $m=0$.

\section{On Nernst's theorem}

It is important to ensure that the present formalism does not come
into conflict with basic thermodynamics. In the present case this
means in particular that the Casimir entropy per unit area at zero
temperature,
\begin{equation}
S=-\left(\frac{\partial F}{\partial T}\right)_V, \label{9}
\end{equation}
($F$ being the free energy), has to be zero at $T=0$. This is
Nernst's theorem. Let us make some brief remarks on this topic, in
view of its current interest.

$\bullet$ There exist no measurements of the permittivities at
very low frequencies. What is at our disposal, is a series of
measurements of room-temperature complex frequencies
$\varepsilon(\omega)=\varepsilon'(\omega)+\varepsilon''(\omega)$,
where the data on $\varepsilon''(\omega)$ permit us to calculate
the real quantities $\varepsilon(i\zeta)$ via the Kramers-Kronig
relation. The permittivity data received from Lambrecht cover the
frequency region $1.5\times 10^{11}$ rad/s to $1.5 \times 10^{18}$
rad/s. Based upon these data, the relaxation frequency $\nu $ in
the Drude relation is determined. For low frequencies we have to
describe the permittivity analytically, with use of the Drude
relation, down to $\zeta =0$. It is not very important, however,
to know the value of $\nu$ at $T=0$ very accurately; the important
point is that $\nu$ at $T=0$ stays finite. In practice, this
condition is always fulfilled because of scattering from
impurities. Then, it is easy to show that the zero frequency TE
mode does not contribute to the Casimir effect. The calculation is
shown explicitly in Appendix A in \cite{hoye03}.

$\bullet$ The above remarks were related to room temperatures. If
we proceed to consider low temperatures, we observe that nor in
this case there exist  permittivity measurements. We thus again
have to take recourse to the Drude relation in which, in
principle, the value of $\nu$ can be different from that above. It
seems, however, that the physical significance of an altered value
of $\nu$ is only minor (a discussion on this point is given in
\cite{brevik05}). This is partly due to the impurities, as
mentioned above, resulting in scattering also when the temperature
is low. Important in the present context is that $\nu$ stays
finite when $T\rightarrow 0$, so that the behaviour is essentially
as above: there is no contribution from the zero frequency TE mode
to the Casimir force. Mathematically, the essential point is that
\begin{equation}
\zeta^2[\varepsilon(i\zeta)-1] \rightarrow 0\quad {\rm as} \quad
\zeta \rightarrow 0. \label{10}
\end{equation}
Correspondingly, if the free energy $F$ is drawn as a function of
$T$ at some fixed value of $a$, it turns out numerically, to a
high precision,  that the slope of the curve is zero at $T=0$.
This result was indicated in figure 5 in \cite{hoye03};  we intend
to deal with the topic in more detail in \cite{hoye06}. That is,
Nernst's theorem is found to be well  satisfied numerically.

$\bullet$ In the above argument the existence of impurities played
a certain role, ensuring that $\nu$ stays different from zero at
all temperatures. Now, it is legitimate to ask: what about the
ideal case where the metal is entirely free from impurities? This
question, although academic, is nevertheless of fundamental
interest.

We may in this context recall the Bloch-Gr{\"u}neisen formula for
the temperature dependence of the electrical resistivity $\rho$
\cite{condon67}. From this one may estimate the temperature
dependence of $\nu=\nu(T)$ to be \cite{brevik05}
\begin{equation}
\nu(T)=0.0847\left(
\frac{T}{\Theta}\right)^5\int_0^{\Theta/T}\frac{x^5e^xdx}{(e^x-1)^2},
\label{11}
\end{equation}
where $\Theta=175\,$K for gold. The Bloch-Gr{\"u}neisen argument
neglects the effect from impurities. It is seen that
$\nu(T)\rightarrow 0$ when $T\rightarrow 0$, so that this case
becomes indeterminate. To deal with this situation, additional
physical effects have to be drawn into  consideration:

1)  One way is to include {\it spatial dispersion}, as was
recently done in \cite{sernelius05}. One finds by this extension
of the theory practically the same results as we did above: there
is only a negligible contribution to the Casimir force from the
zero frequency TE mode. Moreover, the Nernst theorem is found to
be satisfied, so that the presence of dissipation or a finite
relaxation frequency in the material  is not a necessity for this
theorem to hold.

2)  Another approach is to take into account the {\it anomalous
skin effect} \cite{esquivel04,svetovoy05}. This effect is
physically due to the mean free path in the metal being much
larger than the field penetration depth near $T=0$. Again, the
results are found to be essentially the same as above: there is no
contribution to the Casimir force from the zero TE mode, and there
is no contradiction with the Nernst theorem.

\bigskip

 {\bf Acknowledgment}
 \bigskip

 We thank Astrid Lambrecht for
providing us with updated permittivity data.

\newpage
\begin{table}
\begin{tabular}{cccc}

a/$\mu$m         &   T=1 K   &  T=300 K  & T=350 K\\ \hline \hline

 0.16      &   1144      &   1127      &  1124 \\
 0.2       &    508.2    &   497.8     &   495.7     \\
 0.4       &   38.61     &   36.70     &  36.35                             \\
 0.5       & 16.56       &   15.49     &  15.30                            \\
 0.7       &4.556        &4.127        &  4.052                           \\
 1.0       & 1.143       & 0.9852      & 0.9590                           \\
 1.5       & 0.2342      & 0.1856      & 0.1787                             \\
 2.0    & $7.549\times 10^{-2}$  & $5.550\times 10^{-2}$ & $5.344\times 10^{-2}$   \\
 2.5     & $3.128\times 10^{-2}$ & $2.176\times 10^{-2}$  & $2.135\times 10^{-2}$   \\
 3.0     & $1.520\times 10^{-2}$ & $1.033\times 10^{-2}$  & $1.049\times 10^{-2}$   \\
 3.5    & $8.252\times 10^{-3}$  & $5.674\times 10^{-3}$  & $5.990\times 10^{-3}$    \\
 4.0    & $4.858\times 10^{-3}$  & $3.481\times 10^{-3}$  & $3.804\times 10^{-3}$                                          \\
\hline \hline
\end{tabular}
\caption{ The Casimir pressure between Au-Au plates versus gap
width $a$, when $T=\{1, 300, 350\}$ K. The pressures are given in
mPa.}
\end{table}
\begin{table}
\begin{tabular}{cccc}
a/$\mu$m         &   T=1 K   &  T=300 K  & T=350 K\\ \hline \hline
 0.16          &  1141       &  1123       &  1120 \\
 0.2          &   507.4       &  496.8     &   49.47     \\
 0.4           & 38.63       & 36.69       & 36.34    \\
 0.5           & 16.57       & 15.49       & 15.30     \\
 0.7           & 4.560       & 4.127       & 4.052     \\
 1.0           & 1.145       & 0.9854      & 0.9592  \\
 1.5           & 0.2345      & 0.1857      & 0.1787    \\
 2.0           & $7.559\times 10^{-2}$   & $5.551\times 10^{-2}$     & $5.345\times 10^{-2}$ \\
 2.5           & $3.132\times 10^{-2}$   & $2.177\times 10^{-2}$      & $2.135\times 10^{-2}$  \\
 3.0           &$1.522\times10^{-2}$     & $ 1.033\times 10^{-2}$    &$1.049\times 10^{-2}$    \\
 3.5           &$8.263\times 10^{-3}$    &$5.674\times 10^{-3}$     &$5.990\times 10^{-3}$     \\
 4.0           &$ 4.864\times 10^{-3}$     &$3.481\times 10^{-3}$     &$3.805\times 10^{-3}$     \\
\hline \hline
\end{tabular}
\caption{ Same as table 1, but for Cu-Cu plates.}
\end{table}
\begin{table}
\begin{tabular}{cccc}

a/$\mu$m         &   T=1 K   &  T=300 K  & T=350 K\\ \hline \hline

 0.16         & 1290        &   1271      &  1267 \\
 0.2          & 565.3       &   553.9     &   551.6     \\
 0.4          & 41.17       & 39.15       & 38.77      \\
 0.5          & 17.45       & 16.34       & 16.13       \\
 0.7          & 4.734       & 4.290       & 4.212       \\
 1.0          & 1.175       & 1.012       &0.9853         \\
 1.5          & 0.2383      & 0.1889       & 0.1818        \\
 2.0       &$7.648\times 10^{-2}$  &$5.617\times 10^{-2}$ &$5.404\times 10^{-2}$  \\
 2.5       &$3.160\times 10^{-2}$  &$2.195\times 10^{-2}$ &$2.150\times 10^{-2}$  \\
 3.0       &$1.533\times 10^{-2}$  &$1.039\times 10^{-2}$ &$1.053\times 10^{-2}$  \\
 3.5       &$8.311\times 10^{-3}$  &$5.693\times 10^{-3}$ &$ 6.003\times 10^{-3}$  \\
 4.0       &$4.888\times 10^{-3}$  &$3.488\times 10^{-3}$ &$3.809\times 10^{-3}$    \\
\hline \hline
\end{tabular}
\caption{ Same as table 1, but for Al-Al plates.}
\end{table}
\begin{table}
\begin{tabular}{cccc}

a/$\mu$m         &   T=1 K   &  T=300 K  & T=350 K\\ \hline \hline

 0.16       & 1143        & 1125        &  1122 \\
 0.2        & 507.8       & 497.3       &  495.2     \\
 0.4        & 38.62       &36.70        & 36.34      \\
 0.5        & 16.56       &15.49        &15.30       \\
 0.7        & 4.558       & 4.127       &4.052        \\
 1.0        & 1.144       & 0.9853      & 0.9591       \\
 1.5        & 0.2343      & 0.1857      & 0.1787       \\
 2.0    &$ 7.554\times 10^{-2}$ &$ 5.550\times 10^{-2}$ &$ 5.345\times 10^{-2}$  \\
 2.5    &$3.130\times 10^{-2}$  &$ 2.177\times 10^{-2}$ &$ 2.135\times 10^{-2}$ \\
 3.0    &$1.521\times 10^{-2}$  &$1.033\times 10^{-2}$  &$ 1.049\times 10^{-2}$ \\
 3.5    &$8.258\times 10^{-3}$  &$5.674\times 10^{-3}$  &$5.990\times 10^{-3}$    \\
 4.0    &$4.861\times 10^{-3}$  &$3.481\times 10^{-3}$  &$3.805\times 10^{-3}$     \\
\hline \hline
\end{tabular}
\caption{ Same as table 1, but for Au-Cu plates.}
\end{table}
\begin{table}
\begin{tabular}{cccc}

a/$\mu$m         &   T=1 K   &  T=300 K  & T=350 K\\ \hline \hline

 0.16          & 1213        & 1195        & 1191  \\
 0.2          &   535.4      & 524.5       &  522.3  \\
 0.4          & 39.85        & 37.89        &37.52    \\
 0.5          &16.99         &15.90        &15.70     \\
 0.7          &4.643         &4.207       &4.130      \\
 1.0          &1.159         &0.9986      &0.9720     \\
 1.5          &0.2362        &0.1873      &0.1802     \\
 2.0     &$7.598\times 10^{-2}$ &$5.583\times 10^{-2}$ &$5.374\times 10^{-2}$ \\
 2.5     &$3.144\times 10^{-2}$ &$2.185\times 10^{-2}$  &$2.142\times 10^{-2}$ \\
 3.0    &$1.527\times 10^{-2}$ &$1.036\times 10^{-2}$ &$1.051\times 10^{-2}$   \\
 3.5    &$8.281\times 10^{-3}$ &$5.684\times 10^{-3}$ &$5.996\times 10^{-3}$ \\
 4.0    &$4.873\times 10^{-3}$ &$3.485\times 10^{-3}$ &$3.807\times
 10^{-3}$\\
 \hline \hline
\end{tabular}
\caption{ Same as table 1, but for Au-Al plates.}
\end{table}
\begin{table}
\begin{tabular}{cccc}

a/$\mu$m         &   T=1 K   &  T=300 K  & T=350 K\\ \hline \hline

 0.16     &1211     &  1193       &    1189        \\
 0.2      & 535.0   &  524.0      &  521.8      \\
 0.4      &39.86    &37.89        &37.52         \\
 0.5      &17.00     &15.90        &15.70          \\
 0.7      &4.646     &4.207       &4.130           \\
 1.0      &1.159     &0.9987      &0.9720         \\
 1.5      &0.2364     &0.1873      &0.1802         \\
 2.0    &$7.603\times 10^{-2}$ &$5.584\times 10^{-2}$ &$5.375\times 10^{-2}$  \\
 2.5    &$3.146\times 10^{-2}$ &$2.186\times 10^{-2}$ &$2.143\times 10^{-2}$ \\
 3.0    &$1.528\times 10^{-2}$ &$1.036\times 10^{-2}$ &$ 1.051\times 10^{-2}$ \\
 3.5    &$8.287\times 10^{-3}$ &$5.684\times 10^{-3}$ &$5.997\times 10^{-3}$  \\
 4.0     &$4.876\times 10^{-3}$ &$3.485\times 10^{-3}$ &$3.807\times 10^{-3}$  \\
\hline \hline
\end{tabular}
\caption{ Same as table 1, but for Cu-Al plates.}
\end{table}
\end{document}